\batchmode
\documentstyle[12pt]{article}
\newcommand{\nc}{\newcommand}
\nc{\renc}{\renewcommand}

\nc{\half}{{\textstyle{1\over2}}}
\nc{\etal}{\mbox{\it et al. }}
\nc{\ie}{{\it i.e.}}
\nc{\eg}{{\it e.g.}}

\renc{\thefootnote}{\arabic{footnote}}
\nc{\capt}[1]{{\bf Figure.} {\small\sl #1}}

\nc{\eqs}[2]{\mbox{Eqs.~(\ref{#1},\,\ref{#2})}}
\nc{\eq}[1]{\mbox{Eq.~(\ref{#1})}}

\nc{\figs}[2]{\mbox{Figs.~(\ref{#1},\,\ref{#2})}}
\nc{\fig}[1]{\mbox{Fig~.(\ref{#1})}}

\nc{\tag}[1]{\label{#1} \marginpar{{\footnotesize #1}}}
\nc{\mtag}[1]{\label{#1} \mbox{\marginpar{{\footnotesize #1}}}}
\renc{\baselinestretch}{1.5}
\newlength{\overeqskip}
\newlength{\undereqskip}
\nc{\be}[1]{\begin{equation} \mbox{$\label{#1}$}}
\nc{\bea}[1]{\begin{eqnarray} \mbox{$\label{#1}$}}
\nc{\Section}[2]{\section{#2}\label{#1}}
\nc{\Bibitem}[1]{\bibitem{#1}}
\nc{\Label}[1]{\label{#1}}

\nc{\eea}{\vspace{\undereqskip}\end{eqnarray}}
\nc{\ee}{\vspace{\undereqskip}\end{equation}}
\nc{\bdm}{\begin{displaymath}}
\nc{\edm}{\end{displaymath}}
\nc{\dpsty}{\displaystyle}
\nc{\bc}{\begin{center}}
\nc{\ec}{\end{center}}
\nc{\ba}{\begin{array}}
\nc{\ea}{\end{array}}
\nc{\bab}{\begin{abstract}}
\nc{\eab}{\end{abstract}}
\nc{\btab}{\begin{tabular}}
\nc{\etab}{\end{tabular}}
\nc{\bit}{\begin{itemize}}
\nc{\eit}{\end{itemize}}
\nc{\ben}{\begin{enumerate}}
\nc{\een}{\end{enumerate}}
\nc{\bfig}{\begin{figure}}
\nc{\efig}{\end{figure}}
\nc{\arreq}{&\!=\!&}
\nc{\arrmi}{&\!-\!&}
\nc{\arrpl}{&\!+\!&}
\nc{\arrap}{&\!\!\!\approx\!\!\!&}
\nc{\non}{\nonumber\\*}
\nc{\align}{\!\!\!\!\!\!\!\!&&}

\def\lsim{\; \raise0.3ex\hbox{$<$\kern-0.75em
      \raise-1.1ex\hbox{$\sim$}}\; }
\def\gsim{\; \raise0.3ex\hbox{$>$\kern-0.75em
      \raise-1.1ex\hbox{$\sim$}}\; }
\nc{\DOT}{\hspace{-0.08in}{\bf .}\hspace{0.1in}}
\nc{\Laada}{\hbox {$\sqcap$ \kern -1em $\sqcup$}}
\nc\loota{{\scriptstyle\sqcap\kern-0.55em\hbox{$\scriptstyle\sqcup$}}}
\nc\Loota{{\sqcap\kern-0.65em\hbox{$\sqcup$}}}
\nc\laada{\Loota}
\nc{\qed}{\hskip 3em \hbox{\BOX} \vskip 2ex}

\nc{\real}{{\rm I \! R}}
\nc{\Z}{{\sf Z \!\!\! Z}}
\nc{\complex}{{\rm C\!\!\! {\sf I}\,\,}}
\def\bigid{\leavevmode\hbox{\small1\kern-3.8pt\normalsize1}}
\def\id{\leavevmode\hbox{\small1\kern-3.3pt\normalsize1}}
\nc{\slask}{\!\!\!/}
\nc{\bis}{{\prime\prime}}
\nc{\pa}{\partial}
\nc{\na}{\nabla}
\nc{\ra}{\rangle}
\nc{\la}{\langle}
\nc{\goto}{\rightarrow}
\nc{\swap}{\leftrightarrow}

\nc{\EE}[1]{ \mbox{$\cdot10^{#1}$} }
\nc{\abs}[1]{\left|#1\right|}
\nc{\at}[2]{\left.#1\right|_{#2}}
\nc{\norm}[1]{\|#1\|}
\nc{\abscut}[2]{\Abs{#1}_{\scriptscriptstyle#2}}
\nc{\vek}[1]{{\rm\bf #1}}
\nc{\integral}[2]{\int\limits_{#1}^{#2}}
\nc{\inv}[1]{\frac{1}{#1}}
\nc{\dd}[2]{{{\partial #1}\over{\partial #2}}}
\nc{\ddd}[2]{{{{\partial}^2 #1}\over{\partial {#2}^2}}}
\nc{\dddd}[3]{{{{\partial}^2 #1}\over
	{\partial #2 \partial #3}}}
\nc{\dder}[2]{{{d #1}\over{d #2}}}
\nc{\ddder}[2]{{{d^2 #1}\over{d {#2}^2}}}
\nc{\dddder}[3]{{d^2 #1}\over
	{d #2 d #3}}
\nc{\dx}[1]{d\,^{#1}x}
\nc{\dy}[1]{d\,^{#1}y}
\nc{\dz}[1]{d\,^{#1}z}
\nc{\dl}[1]{\frac{d\,^{#1}l}{(2\pi)^{#1}}}
\nc{\dk}[1]{\frac{d\,^{#1}k}{(2\pi)^{#1}}}
\nc{\dq}[1]{\frac{d\,^{#1}q}{(2\pi)^{#1}}}

\nc{\cc}{\mbox{$c.c.$ }}
\nc{\hc}{\mbox{$h.c.$ }}
\nc{\cf}{cf.\ }
\nc{\erfc}{{\rm erfc}}
\nc{\Tr}{{\rm Tr\,}}
\nc{\tr}{{\rm tr\,}}
\nc{\pol}{{\rm pol}}
\nc{\sign}{{\rm sign}}
\nc{\bfT}{{\bf T }}

\def\GeV{{\rm\ GeV}}

\nc{\cA}{{\cal A}}
\nc{\cB}{{\cal B}}
\nc{\cD}{{\cal D}}
\nc{\cE}{{\cal E}}
\nc{\cG}{{\cal G}}
\nc{\cH}{{\cal H}}
\nc{\cL}{{\cal L}}
\nc{\cO}{{\cal O}}
\nc{\cT}{{\cal T}}
\nc{\cN}{{\cal N}}
\nc{\rvac}[1]{|{\cal O}#1\rangle}
\nc{\lvac}[1]{\langle{\cal O}#1|}
\nc{\rvacb}[1]{|{\cal O}_\beta #1\rangle}
\nc{\lvacb}[1]{\langle{\cal O}_\beta #1 |}
\nc{\bb}{\bar{\beta}}
\nc{\bt}{\tilde{\beta}}
\nc{\ctH}{\tilde{\cal H}}
\nc{\chH}{\hat{\cal H}}
\nc{\1}{\aa}
\nc{\2}{\"{a}}
\nc{\3}{\"{o}}
\nc{\4}{\AA}
\nc{\5}{\"{A}}
\nc{\6}{\"{O}}
\nc{\al}{\alpha}
\nc{\g}{\gamma}
\nc{\Del}{\Delta}
\nc{\e}{\epsilon}
\nc{\eps}{\epsilon}
\nc{\lam}{\lambda}
\nc{\om}{\omega}
\nc{\Om}{\Omega}
\nc{\ve}{\varepsilon}
\nc{\mn}{{\mu\nu}}
\nc{\k}{\kappa}
\nc{\vp}{\varphi}

\nc{\advp}[3]{{\it  Adv.\ in\ Phys.\ }{{\bf #1} {(#2)} {#3}}}
\nc{\annp}[3]{{\it  Ann.\ Phys.\ (N.Y.)\ }{{\bf #1} {(#2)} {#3}}}
\nc{\apl}[3]{{\it  Appl. Phys. Lett. }{{\bf #1} {(#2)} {#3}}}
\nc{\apj}[3]{{\it  Ap.\ J.\ }{{\bf #1} {(#2)} {#3}}}
\nc{\apjl}[3]{{\it  Ap.\ J.\ Lett.\ }{{\bf #1} {(#2)} {#3}}}
\nc{\app}[3]{{\it Astropart.\ Phys.\ }{{\bf #1} {(#2)} {#3}}}
\nc{\cmp}[3]{{\it  Comm.\ Math.\ Phys.\ }{{ \bf #1} {(#2)} {#3}}}
\nc{\cqg}[3]{{\it  Class.\ Quant.\ Grav.\ }{{\bf #1} {(#2)} {#3}}}
\nc{\epl}[3]{{\it  Europhys.\ Lett.\ }{{\bf #1} {(#2)} {#3}}}
\nc{\ijmp}[3]{{\it Int.\ J.\ Mod.\ Phys.\ }{{\bf #1} {(#2)} {#3}}}
\nc{\ijtp}[3]{{\it Int.\ J.\ Theor.\ Phys.\ }{{\bf #1} {(#2)} {#3}}}
\nc{\jmp}[3]{{\it  J.\ Math.\ Phys.\ }{{ \bf #1} {(#2)} {#3}}}
\nc{\jpa}[3]{{\it  J.\ Phys.\ A\ }{{\bf #1} {(#2)} {#3}}}
\nc{\jpc}[3]{{\it  J.\ Phys.\ C\ }{{\bf #1} {(#2)} {#3}}}
\nc{\jap}[3]{{\it J.\ Appl.\ Phys.\ }{{\bf #1} {(#2)} {#3}}}
\nc{\jpsj}[3]{{\it J.\ Phys.\ Soc.\ Japan\ }{{\bf #1} {(#2)} {#3}}}
\nc{\lmp}[3]{{\it Lett.\ Math.\ Phys.\ }{{\bf #1} {(#2)} {#3}}}
\nc{\mpl}[3]{{\it  Mod.\ Phys.\ Lett.\ }{{\bf #1} {(#2)} {#3}}}
\nc{\ncim}[3]{{\it  Nuov.\ Cim.\ }{{\bf #1} {(#2)} {#3}}}
\nc{\np}[3]{{\it  Nucl.\ Phys.\ }{{\bf #1} {(#2)} {#3}}}
\nc{\npps}[3]{{\it  Nucl.\ Phys.\ Proc.\ Suppl.\ }{{\bf #1} {(#2)} {#3}}}
\nc{\pr}[3]{{\it Phys.\ Rev.\ }{{\bf #1} {(#2)} {#3}}}
\nc{\pra}[3]{{\it  Phys.\ Rev.\ A\ }{{\bf #1} {(#2)} {#3}}}
\nc{\prb}[3]{{\it  Phys.\ Rev.\ B\ }{{{\bf #1} {(#2)} {#3}}}}
\nc{\prc}[3]{{\it  Phys.\ Rev.\ C\ }{{\bf #1} {(#2)} {#3}}}
\nc{\prd}[3]{{\it  Phys.\ Rev.\ D\ }{{\bf #1} {(#2)} {#3}}}
\nc{\prl}[3]{{\it Phys.\ Rev.\ Lett.\ }{{\bf #1} {(#2)} {#3}}}
\nc{\pl}[3]{{\it  Phys.\ Lett.\ }{{\bf #1} {(#2)} {#3}}}
\nc{\prep}[3]{{\it Phys.\ Rep.\ }{{\bf #1} {(#2)} {#3}}}
\nc{\prsl}[3]{{\it Proc.\ R.\ Soc.\ London\ }{{\bf #1} {(#2)} {#3}}}
\nc{\ptp}[3]{{\it  Prog.\ Theor.\ Phys.\ }{{\bf #1} {(#2)} {#3}}}
\nc{\ptps}[3]{{\it  Prog\ Theor.\ Phys.\ suppl.\ }{{\bf #1} {(#2)} {#3}}}
\nc{\physa}[3]{{\it  Physica\ A\ }{{\bf #1} {(#2)} {#3}}}
\nc{\physb}[3]{{\it  Physica\ B\ }{{\bf #1} {(#2)} {#3}}}
\nc{\phys}[3]{{\it Physica\ }{{\bf #1} {(#2)} {#3}}}
\nc{\rmp}[3]{{\it  Rev.\ Mod.\ Phys.\ }{{\bf #1} {(#2)} {#3}}}
\nc{\rpp}[3]{{\it Rep.\ Prog.\ Phys.\ }{{\bf #1} {(#2)} {#3}}}
\nc{\sjnp}[3]{{\it Sov.\ J.\ Nucl.\ Phys.\ }{{\bf #1} {(#2)} {#3}}}
\nc{\spjetp}[3]{{\it Sov.\ Phys.\ JETP\ }{{\bf #1} {(#2)} {#3}}}
\nc{\yf}[3]{{\it Yad.\ Fiz.\ }{{\bf #1} {(#2)} {#3}}}
\nc{\zetp}[3]{{\it Zh.\ Eksp.\ Teor.\ Fiz.\  }{{\bf #1}  {(#2)} {#3}}}
\nc{\zp}[3]{{\it Z.\ Phys.\ }{{\bf #1} {(#2)} {#3}}}
\nc{\ibid}[3]{{\sl ibid.\ }{{\bf #1} {#2} {#3}}}
\nc{\rf}[1]{(\ref{#1})}
\nc{\nn}{\nonumber \\*}
\nc{\bfB}{\bf{B}}
\nc{\bfv}{\bf{v}}
\nc{\bfx}{\bf{x}}
\nc{\bfy}{\bf{y}}
\nc{\vx}{\vec{x}}
\nc{\vy}{\vec{y}}
\nc{\oB}{\overline{B}}
\nc{\oI}{\overline{I}}
\nc{\oR}{\overline{R}}
\nc{\rar}{\rightarrow}
\nc{\ti}{\times}
\nc{\slsh}{\hskip-5pt/}
\nc{\sm}{Standard~Model~}
\nc{\MP}{M_{\rm Pl}}
\nc{\tp}{t_{\rm Pl}}
\nc{\ave}{\bar{E}}

\nc{\eff}{{\rm eff}}
\nc{\kk}{\vek{k}}
\nc{\pp}{{\rm p}}
\nc{\ga}{g_{a\gamma}}
\nc{\vv}{\\}
\nc{\eee}{{\bf E}}
\nc{\bbb}{{\bf B}}
\nc{\qcd}{T_{\rm QCD}}
\nc{\G}{\rm \ G}
\def\vec#1{{\bf #1}}

\def\lae{\;^{<}_{\sim} \;} \def\gae{\; ^{>}_{\sim} \;} 

\def\udd{u^{c}d^{c}d^{c}}

\begin{document}
{\title{\vskip-2truecm{\hfill {{\small \\
	\hfill GUTPA/99/XX/X \\
	}}\vskip 1truecm}
{\bf  B-ball Baryogenesis and D-term Inflation$^*$}}
{\author{
{\sc  John McDonald$^{1}$}\\
{\sl\small Department of Physics and Astronomy, University of Glasgow,
Glasgow G12 8QQ, SCOTLAND}
}
\maketitle
\begin{abstract}
\noindent
              
                The MSSM has flat directions in its scalar potential, along which it is natural 
for Bose condensates of squarks to form in the early Universe.
A baryon asymmetry can be induced in these condensates via Affleck-Dine baryogenesis. The condensates are unstable with respect to fragmentation to "B-balls", solitons made of squarks and carrying baryon number, 
which, if they survive thermalization, fill the Universe down to low temperatures, much lower than that of the electroweak phase transition, with interesting cosmological consequences. 
In particular, their decay implies a similar number density of baryons and dark matter neutralinos, 
in accordance with observations. Evasion of thermalization and the
ability to account for the observed baryon asymmetry requires a very low reheating temperature, 
which, it is argued, is a natural feature of presently favoured D-term inflation models. 
\end{abstract}
\vfil
\footnoterule
{\small $^*$Presented at "Strong and Electroweak Matter '98",
NORDITA, Copenhagen}
\\
{\small $^1$mcdonald@physics.gla.ac.uk}

\thispagestyle{empty}
\newpage
\setcounter{page}{1}


     The possibility of electroweak baryogenesis \cite{ewb}
 in the Minimal SUSY Standard Model (MSSM) \cite{nilles} has become increasingly tightly constrained by
experiment. In particular, the requirement that the 
electroweak phase transition is strong enough to prevent
subsequent wash-out of the
asymmetry imposes an upper bound on the Higgs
mass of 105-107 GeV \cite{Hupper} (the present 95 $\%$ CL LEP
lower bound is 77.5GeV \cite{Hlower},
with almost all of the remaining range to be tested by LEP200 \cite{Hupper}), and in addition requires that the right-hand 
stop is light \cite{lights}, which in turn requires a negative SUSY breaking mass squared term for the r.h. stop 
(but not for the other squarks and sleptons); a stiff challenge for supergravity (SUGRA) models.
 However, there is another perfectly natural way to generate the baryon asymmetry 
in the context of the MSSM, namely Affleck-Dine baryogenesis \cite{ad}. In this the 
B asymmetry is induced in a squark Bose condensate which subsequently decays to (or is thermalized to) quarks.

    This mechanism has recently been given a new spin \cite{kus,ks,bbb1} with the realization that in almost all realistic cases the squark condensate is unstable with respect to 
fragmentation to B-balls (B carrying Q-balls \cite{Qballs,Qballs2}) made of squarks, which fill the Universe for a significant part of its history (until well after the electroweak phase transition in the case of gravity-mediated SUSY breaking 
\cite{bbb1,bbb2} and right up to and including NOW, with interesting experimental and astrophysical consequences, in the case of gauge-mediated SUSY breaking \cite{ks,ks234}). These B-balls can have very significant consequences for cosmology;
in particular, in the case of
gravity-mediated SUSY breaking on which
we focus here, they can explain the
remarkable similarity of the {\it number density}
of baryons and dark matter particles for
the case where the dark matter
particles have masses $O(m_W)$ \cite{bbb2,bbbdm},
such as is typically the case for neutralinos.
To be precise, with currently favoured values compatible with distant supernovae observations \cite{sn},
 $\Omega = 1$, $\Omega_{m} = 0.4$, and $\Omega_{\Lambda} = 0.6$, expansion rate $h$ in the range 0.6 to 0.8 (as determined by Hubble Space Telescope 
\cite{hst,freedman} and limits on the age of the Universe \cite{ageu}) and 
baryon number density $0.0048 \lae \Omega_{B}h^2 \lae 0.013$, as determined by primordial
nucleosynthesis \cite{sarkar}, we find
\be{n1} \frac{n_{B}}{n_{DM}} \approx (1.5-7.3) \; \frac{m_{DM}}{m_W}     ~. \ee
Thus for $m_{DM} \sim m_W$, the baryon and dark matter particle number densities 
are within an order of magnitude of each other; a remarkable result and a possible clue as to the
origin of baryons and dark matter. Although one might expect that the existence of 
a significant mass density of baryons could be understood by appealing to anthropic selection, there
is still no reason for such a selection to
result in similar {\it number} densities
of baryons and dark matter. We refer to the scenario in which the B asymmetry at least partly comes from late-decaying B-balls as B-ball baryogenesis.

 The details of this mechanism are intimately tied to the details of the full 
supergravity (SUGRA) model and inflation, requiring a rather
 global view of SUSY cosmology. This connection with SUSY cosmology as a whole, and the possibility of observable predictions \cite{bbbdm,bbbiso} (as discussed in 
more detail in these Proceedings by Enqvist), 
make this an exciting scenario, requiring, we emphasize, nothing more than the particles of the MSSM and a model for inflation.

           B-ball baryogenesis is based on the
           Affleck-Dine mechanism. This in turn depends on the
           existence of renormalizable flat directions in the
           scalar potential of the MSSM \cite{drt}. The scalar potential of the MSSM is complicated, with squarks and sleptons as well as Higgs fields. Renormalizable flat directions correspond to directions in the scalar potential with renormalizible F- and D-terms
vanishing, leaving only the soft SUSY-breaking terms and the Planck-suppressed 
non-renormalizable terms which are expected in a SUGRA effective theory. These directions are characterized by the lowest dimension superpotential operators which are consistent with all symmetries and which which are non-vanishing when expressed as scalars along the flat directions. 
We concentrate on the case of the MSSM with R-parity, which eliminates dangerous renormalizable B- and L-violating terms from
the superpotential and also allows for stable neutralino cold dark
matter \cite{nilles,susydm}. In this case
the flat directions are of even dimension \cite{drt} e.g.
\newline d=4 $H_{u}L$-direction
\be{e1} <\tilde{\nu}_L> = < \phi_{u}^{o}> \;\;\; ; \;\;\; (H_{u}L)^2 \neq 0  ~,\ee
\newline d=6 $u^c d^c d^c$-direction (the $d^cQL$ and $e^cLL$ directions are similar)
\be{e2} <\tilde{u}^{c\;1}> = <\tilde{d}^{c\;2}>  = <\tilde{d}^{c'\;3}>   \;\;\; ; \;\;\; (\udd)^2 \neq 0  ~,\ee
where the indices denote colour and $\tilde{d}^{c}$ and $\tilde{d}^{c'}$ are orthogonal in flavour
space. We will be particularly interested in the d=6 $\udd$ direction, as d=4 directions are both incompatible with 
B-balls which survive thermalization \cite{bbb2} and with the AD mechanism in the context of 
D-term inflation \cite{kmr}.

     The form of the soft SUSY-breaking terms (once Planck-suppressed SUGRA
couplings are taken into account) is different in the early Universe; the 
energy density of the Universe breaks SUSY and results in O(H) corrections 
to the SUSY-breaking terms \cite{h2O,h2}. As a result, it is natural for the
initial value of the AD squark field to take a large value during and/or soon
after inflation. This is true if the correction to the SUSY breaking mass squared term is 
negative. The scalar potential then has the form 
\be{e3} V = (m^2 - c H^2) |\Phi|^2 + \left( \frac{A_{\lambda} \lambda \Phi^6}{M^3} + h.c.\right) +
\frac{\lambda^2 |\Phi|^{10}}{M^6}    ~,\ee
where $M = M_{Pl}/\sqrt{8 \pi}$ and $c \sim 1$. 
The AD field will begin to oscillate coherently once $H \approx m$, 
corresponding to a Bose condensate of squarks, with an initial amplitude 
$\phi \approx (m M^3)^{1/4}\approx 10^{14} \GeV$ for the d=6 direction and 
$\phi \approx (m M)^{1/2} \approx 10^{10} \GeV$ along
the d=4 direction. (It is the larger amplitude
along the d=6 direction which protects the
AD field from being thermalized
in D-term inflation models, by giving a large mass to the particles which couple to it, so
suppressing the scattering rate with light thermal particles \cite{kmr}). 
In the absence of the A-terms, there is no B violation
or CP violation, and the real and imaginary parts of the AD scalar
oscillate in phase, corresponding to zero B asymmetry. 
A B asymmetry is induced in the squark condensate if a phase shift is induced between the real and
imaginary parts of the AD field. The baryon asymmetry density is given by
\be{e4} n_{B} = i( \dot{\Phi}^{\dagger}\Phi - \Phi^{\dagger} \dot{\Phi})   ~.\ee
With $\Phi = (\phi_1 + i \phi_2)/\sqrt{2}$, $\phi_1 = \phi_o Cos(mt)$ and $\phi_2 = \phi_o Cos(mt + \delta)$, 
the B asymmetry density is given by $n_B = (m \phi_o^2 Sin \delta )/2 $. The required 
phase shift is provided by the A-term. 
When the AD field starts oscillating, the A-term is of the same order of magnitude as the  
mass term and non-renormalizable term in the potential, and distinguishes between the real and 
imaginary directions (where the real direction is defined by the phase of the A-term). For the case of interest to us here, 
D-term inflation \cite{dti}, in which there are no $O(H)$ corrections to the A-terms \cite{kmr}, the initial phase of the AD field is determined by its random initial value
during inflation and is therefore typically of the order of 1; thus within a few oscillations a phase difference of 
order 1 will be induced in the AD field. Subsequently, as the Universe expands, the amplitude of oscillation
decreases; the A-term, which is proportional to $\phi^6$, rapidly decreases relative to the mass term, 
effectively switching off B violation and leaving a fixed B asymmetry which we observe today. The final 
B-asymmetry depends crucially on the reheating temperature after inflation; for the d=6 $\udd$ direction 
the baryon to entropy ratio is \cite{bbbdti}
\be{e5} \eta_B \approx 3 \times 10^{-11} \left(\frac{T_R}{1 \GeV}\right) Sin \delta    ~.\ee
Comparing with the observed B asymmetry, $\eta_{B\;obs} = (3-8) \times 10^{-11}$, we see that the reheating
temperature for d=6 AD baryogenesis must of the order of 1GeV in
 D-term inflation models (or in any model in which the magnitude of the CP violating phase
is of the order of 1; a natural value); for higher
reheating temperatures there is {\it overproduction} of B. 

           Thus d=6 AD baryogenesis with O(1) CP violating phase (as in the D-term inflation scenario)
imposes the constraint on inflation models that the reheating temperature must be very low relative to the 
energy scale of inflation. At first sight this seems 
perhaps unfavourable; however, reheating in SUSY inflation models must already occur at a low temperature    
relative to the inflation energy scale, $T_R \lae 10^9 GeV$, in order to avoid thermally overproducing 
gravitinos \cite{sarkar,grav}. We will see that there is good reason to expect a very low reheating temperature in
 D-term inflation models, resulting in an interesting self-consistency of the D-term inflation/AD baryogenesis scenario \cite{bbbdti}.      

     In the original AD mechanism it was assumed that the AD field would simply oscillate around the
minimum of its potential with decreasing amplitude until it either decayed or the AD field strength was small enough to allow it to thermalize.  
However, it was recently realized that the AD condensate is generally unstable with respect to 
fragmentation to B-balls \cite{ks,bbb1}. 
In general, Q-ball solutions (charged non-topological solitons in a scalar theory with an approximate
$U(1)_{global}$, which in our case corresponds to baryon number) exist if $U(|\phi|)/|\phi|^2$ has a global minimum at $\phi \neq 0$ \cite{Qballs}. 
This is likely to occur rather generally in SUSY models, 
as a result of the soft SUSY breaking terms \cite{kus}.  
The resulting solution has the form \cite{Qballs} 
\be{e6} \Phi(r,t) = \frac{e^{i \omega t}}{\sqrt{2}} \phi(r)   \;\;\; ; \;\;\; \phi(r) \approx \phi(0) e^{-r^2/R^2}  ~,\ee
where $\omega$ is the effective mass of the scalars inside the Q-ball and
$\phi(r)$ here corresponds to the case of thick-walled
Q-balls \cite{bbb2}. For the particular case of flat directions
in models with gravity-mediated SUSY breaking, the mass squared
term rather generally becomes
smaller with increasing $\phi$ once radiative corrections are taken into account, as a result 
of gauge loops \cite{bbb1} \begin{footnote}{Let us note some interesing work in connection with flat directions, proton stability, Q-balls and string models in this context in \cite{minos}.}\end{footnote}  The RG equations generally have the form \cite{nilles}
\be{e7} \mu \frac{\partial m_{i}}{\partial \mu} = \alpha_{i} m_{i}^{2} - \beta_{\alpha} M_{\alpha}^{2} 
~,\ee 
where $m_{i}$ represents the scalar soft SUSY breaking terms and
$M_{\alpha}$ the gaugino masses. 
These are typically
dominated by the gaugino masses,
causing $m_{i}$ to decrease with increasing scale.
Thus the AD scalar mass term decreases with increasing $\phi$. As a result, 
the condition for the existence of Q-ball solutions is satisfied. 
The homogeneous condensate 
will naturally fragment to Q-balls, as the
Q-ball is the lowest energy solution for a given charge.
This can be also seen by noting that a homogeneous scalar field oscillating in a
potential that is flatter than
$\phi^2$ will behave as matter with a negative pressure \cite{turner}.
At 1-loop the correction to the mass term is 
\be{e8} V(|\Phi|)  = m^2 |\Phi|^2
+... \rightarrow m^2 \left(1 + K log \left(\frac{|\Phi|^2}{\mu^2}\right)\right) |\Phi|^2
+...   ~,\ee
where typically $|K| \approx 0.01 -0.1$ and $ K < 0$.
For small $|K|$, $V \sim \phi^{2+K}$, which results in
an equation of state for the homogeneous field
\be{e9} p = \frac{K}{2} \rho ~.\ee
This negative pressure merely reflects the attractive interaction
between the scalars due to the $log$ term in the potential.
As a result, any spatial perturbations
will grow exponentially with time until they go non-linear and the
condensate will fragment into charged condensate lumps
which eventually condense into B-balls. The
exponential growth of the linear perturbations
is given by \cite{bbb2} 
\be{e10} \delta \phi = \left(\frac{a_{o}}{a}\right)^{3/2}
\delta \phi_{o} exp\left(\frac{2}{H} \left(\frac{|K|}{2} \frac{\vec{k}^2}{a^2}
\right)^{1/2} \right)           ~,\ee
where  $\delta \phi_{o}$ is the
seed perturbation, expected to come from quantum fluctuations during
inflation \cite{bbb1}.
This is valid for perturbations with $|\vec{k}^2/a^2| \lae |2Km^2|$;
shorter wavelength perturbations have a positive
pressure due to their gradient energy which overcomes the negative pressure,
such that the perturbations merely oscillate.
The first perturbation to go non-linear has a diameter
$ l_{non-linear} \approx  \pi / (2 |K| m^2)^{1/2}$.
The baryon number density at a given value of $H$
is given by $n_{B} = \eta_{B} H^2 M_{Pl}^2 / 2 \pi T_R$. 
From these we find that the baryon number of
the condensate lumps (and so of the B-balls) is \cite{bbb2}
\be{e11} B \approx 10^{24} f_B |K|^{1/2}
 \left( \frac{1 \GeV}{T_R} \right)       ~,\ee
with the corresponding radius and field inside the thick-walled B-ball given by
 $R \approx (|K|^{1/2} m)^{-1}$ and
$\phi(0) \approx 10^{14} \left(B/10^{26}\right)^{1/2} \GeV$.
Here $f_B$ is fraction of the B asymmetry which
ends up inside the B-balls; this requires
an analysis of the non-linear evolution of the condensate \cite{fb}
but we expect B-ball formation to be quite efficient; $f_B \sim 0.1-1$. 

          Due to their large charges and field strengths, the B-balls can survive thermalization
and decay at low temperatures. It may be shown that they survive thermalization 
if $T_{R} \lae 10^{3-5}\GeV$ \cite{bbb2}. This is
easily satified for the d=6 D-term inflation case. (However, it rules out
survival of B-balls for the d=4 directions,
since in this case $T_{R} \gae 10^7 \GeV$ is required to
account for the observed B \cite{bbb2}).
Charge escapes from B-balls to light B-carrying fermions and
scalars a rate proportional to their area, $A$.
The decay rate of the B-balls is given by \cite{Qballs2,bbb2}
\be{e12}   \frac{1}{B} \frac{dB}{dt}
\approx - \frac{\omega^3 f_{s}}{192 \pi^2} \frac{A}{B} \propto \frac{1}{B}
   ~,\ee
 where $\omega \approx (1 + 3K/2)m \approx m$, with $m \sim 100GeV$ being the
 mass of the AD scalar \cite{bbb2}.
($f_s$ is a factor accounting for the enhancement of the decay rate if
decay to scalar pairs is possible; we estimate $f_s \approx 10^3$ \cite{bbb2}). 
Therefore larger charge B-balls decay later. The B-balls decay once
 $ | \frac{1}{B} \frac{dB}{dt}| \gae H$; the resulting decay
temperature is 
\be{e13} T_{d} \approx (0.01-1) \left(\frac{T_{R}}{1 \GeV}\right)^{1/2} \GeV           ~.\ee 
Thus, in general, B-balls which can evade thermalization will
decay after the electroweak phase transition.
This implies that the B-balls will have non-trivial
consequences for cosmology:
\newline $\bullet$ The baryon asymmetry inside the B-balls (where thermal particles cannot penetrate)
will be protected from washout due the combined effects of sphaleron B+L violation and additional L violating interactions. Such interactions commonly
arise in extensions of the MSSM; for example, in see-saw models of Majorana neutrino masses. 
\newline $\bullet$ $n_{B} \sim n_{DM}$: If the reheating temperature is of the order of 1GeV, 
then the B-balls will decay {\it after the dark matter neutralinos have frozen out of chemical equilibrium}. Neutralinos freeze out at $T_{f} \approx m_{\chi}/20 > 1 \GeV$, where $m_{\chi}$
is the neutralino mass \cite{susydm}. 
The B-ball is made of squarks, with one unit of R-parity per 1/3 B. Thus when the B-ball decays we will obtain 
3 neutralino LSPs per unit B.
This naturally results in $n_{B} \sim n_{DM}$, in accordance with
observations! Note that this produces {\it non-thermal}
neutralino dark matter, with a quite different dependence on the
MSSM parameter space than the conventional thermal relic dark matter,
a feature that can be tested experimentally
once SUSY is discovered in accelerators.
The actual number density ratio predicted by B-ball decay
(assuming no subsequent annihilations \cite{bbb2}) is crucially dependent
on $f_B$
\be{e15} \frac{n_{B}}{n_{\chi}} = \frac{1}{3 f_{B}}      ~.\ee
Thus $f_{B}$ has to be less than 1 in order to be
compatible with observations (Eq.(1)) and the present
experimental lower bound on the LSP neutralino mass, $m_{\chi} \gae
25GeV$ \cite{mneut}.
However, $f_B$ in the range 0.1 to 1 can accomodate a
wide range of neutralino masses. In addition, if we can {\it calculate} $f_{B}$,
we can constrain the LSP mass. For example, for the $\Omega_{m} = 0.4$ case we find
\be{e15a} 0.046 f_B^{-1} \lae \frac{m_{\chi}}{m_{W}} \lae 0.22 f_B^{-1}
~,\ee
where the range is due to the uncertainties in the baryon asymmetry
estimate from nucleosynthesis and in the present expansion rate.
Since $\Omega_{m}$, $\Omega_{B}$ and $h$ will all be fixed to 1 $\%$ accuracy
by the MAP and PLANCK satellite missions \cite{mapp,freedman}, we will eventually be able to
{\it predict} $m_{\chi}$ if we can calculate $f_{B}$,
providing a "smoking-gun"
test of this scenario \cite{bbbdm}. It remains to be seen whether $f_{B}$ can be accurately
computed from the non-linear evolution of the condensate \cite{fb}. 
\newline $\bullet$ Another observable consequence of late-decaying
B-balls, in the context of D-term inflation models, is that they
transfer the isocurvature fluctuations in the baryons
to the dark matter neutralinos, resulting in a significant 
enhancement of the isocurvature fluctuations,
making them observable by MAP and PLANCK \cite{bbbiso}.
This is discussed in detail in these Proceedings by Enqvist.

      So far we have shown that the d=6 Affleck-Dine
mechanism with O(1) CP violating
phase requires a reheating temperature of the order of 1GeV,
and we have noted that
O(1) phases are to be expected in D-term inflation
models. We now
address the question of whether such low reheating temperatures are a natural feature of these models. 

   SUSY inflation is characterized by the nature of the energy density driving inflation. 
The SUSY potential has F- and D-term contributions,
$V_{susy} = |F|^2 + |D|^2$, with 
D-term inflation driven by $|D|^2$. The reason D-term inflation
is favoured over F-term is that there are no O(H) corrections to the inflaton mass in this
case \cite{dti}. These terms would
otherwise ruin the flatness of the inflaton potential, preventing
slow-rolling and/or producing too large deviations from a scale-invariant
perturbation spectrum.

       The minimal model for D-term inflation has a singlet $S$ and two fields with opposite
$U(1)_{FI}$ charges $\psi_{\pm}$, where $U(1)_{FI}$ is the Fayet-Illiopoulos (FI) gauge group with a FI 
term $\xi^2$. The superpotential is 
\be{e16} W = \lambda S \psi_{+} \psi_{-}  ~\ee
and the scalar potential, including the $U(1)_{FI}$ D-term, is then
\be{e17} V = |\lambda|^2 (|\psi_{+} \psi_{-} |^2 + |S \psi_{+}|^2 + |S \psi_{-}|^2) 
+ \frac{g^2}{2} ( |\psi_{+}|^2 - |\psi_{-}|^2 + \xi^2)^2      ~,\ee
where $g$ is the $U(1)_{FI}$ coupling. $\xi$ is fixed by cosmic microwave background measurements 
to be $6.6 \times 10^{15} \GeV$ \cite{lr}. The inflation
model which results from this potential has the form of a
hybrid inflation model \cite{hybrid}; for
$S > S_{crit} \equiv g \xi/\lambda$ the minimum of the
potential corresponds to $ \psi_{\pm} = 0$,
resulting in a very flat, slow-roling potential  
with $V = g^2 \xi^4 / 2 + \kappa \; log|S|$, where $\kappa$
comes from radiative corrections.
For sufficient inflation, $S > S_{55} \approx 0.9 M$
is necessary. However, this leads to a
generic problem for D-term inflation models \cite{kmr}.
In general, since we are considering SUSY models
to be the low energy effective theory of a SUGRA theory,
we expect all non-renormalizable
superpotential terms consistent with the
symmetries and suppressed by powers of $M$ to appear,
$\Delta W \sim S^m / M^{m-3}$.
However, the contribution to the scalar potential of such terms
will cause unacceptable deviation from
scale-invariant fluctuations at length scales corresponding to the observable Universe,
requiring the elimination of all such terms
for $ m \leq 9$ \cite{kmr}.
Thus we need to introduce a symmetry to eliminate such terms; we focus
on the case of an R-symmetry, which is
particularly effective (eliminating {\it all} purely $S$
superpotential terms if R(S) is negative,
for example), although we expect the same argument to apply
for more general discrete symmetries. The
connection between D-term inflation and low reheating
temperatures is then that the
symmetry which keeps the inflaton potential sufficiently flat
also tends to suppress the coupling of the
inflaton to light fields, so suppressing the decay rate of the
inflaton and resulting in
very low reheating temperatures \cite{bbbdti}
\begin{footnote}{ In D-term inflation models there are {\it two} stages
of reheating; that from decay of the $\psi_{-}$ oscillations, with
typically $T_{R}^{\psi} \approx 10^{15} \GeV$, and that from decay of the
$S$ oscillations \cite{kmr}. The first stage of reheating is important for
thermalizing the d=4 AD field, but otherwise the important reheating
temperature (for the gravitino bound and for B-balls) is that associated
with the $S$ field, whose oscillations dominate the energy
density throughout \cite{bbbdti}.}\end{footnote}.
This happens if there is a "mismatch" between the
transformation properties of the
inflaton sector $S, \psi_{\pm}$ and the MSSM sector fields.
As a simple example, consider the
case where the inflaton sector fields
have half-integral charges and the MSSM
sector fields have 1/3-integral charges.
With $R(S) = -n$, $R(\psi_{+}) = R(\psi_{-}) = (n+2)/2$
($n \in Z$) for the inflaton sector fields and
$R(Q) = R(L) = 5/3$, $R(u^c) = R(d^c) = R(e^c) = 1/3$
and $R(H_u) = R(H_d) = 0$ for the
MSSM fields (the MSSM charges have been chosen to allow
the $(\udd)^2$ operator necessary for AD baryogenesis)
we find that, for n=1, the lowest dimensional superpotential term
allowing the inflaton to couple to the MSSM sector is of the form $W_{int} 
= \kappa S(LH_{u})(u^c u^c d^c e^c)/M^{4}
\sim \kappa S \phi^r / M^{r-2}$ with r=6, where $\phi$ represents the light MSSM
fields \cite{bbbdti}. For n=2 a dimension r = 4 operator is
possible, but for larger n the dimension is $r \geq 6$.
The corresponding decay rate
is 
\be{e18} \Gamma_{S} \approx \left( \frac{M_{S}}{M} \right)^{2(r-2)} \kappa^2 \beta_r M_S 
   ~,\ee   
where $\beta_r$ represents the phase factor for decay to $r$ light particles. 
The corresponding reheating temperature is then 
\be{e19}  T_{R}^S \approx 4 \times 10^{15} (2.8 \times 10^-3)^{r-5/2} \kappa \lambda^{r-3/2} \beta_{r}^{1/2} \GeV  ~.\ee      
Thus for r=6 we find \cite{bbbdti}
\be{e19a} T_{R}^{S} \approx
150 \left( \frac{\lambda}{0.1} \right)^{9/2}
\kappa \beta_{6}^{1/2} \GeV   ~,\ee
where $\beta_{6} \approx 10^{-6}$. Therefore low reheating temperatures,
as low as 1 GeV or less,
are a natural feature of D-term inflation models \cite{bbbdti}. 

     Of course, it is also necessary to check if reheating mechanisms other than single particle decay could be more efficient; in particular parametric resonance \cite{pr} via 
Planck-suppressed operators. We find that parametric resonance is completely negligible if  $ \lambda/g^{1/2} \gae 0.03$ \cite{bbbdti}. For values of $\lambda$ and $g$ not too small compared with 1 this condition will be easily satisfied.

        In conclusion, we believe that Affleck-Dine baryogenesis with late
        decaying B-balls (B-ball baryogenesis)
is a natural candidate for baryogenesis in the MSSM which has
the great advantage that it can explain why $n_B \sim n_{DM}$.
This model fits in very well with the currently favoured D-term inflation 
scenario, with a natural consistency between the low
reheating temperature required for
d=6 AD baryogenesis and the suppression of 
inflaton couplings required for a flat inflaton potential.
It is especially interesting
that this model can make definite predictions;
$n_B \sim n_{DM}$, non-thermal neutralino
dark matter, eventually (depending on the calculation of $f_B$ and results from MAP and PLANCK) the neutralino mass, and,
as discussed in these Proceedings by
Enqvist, isocurvature microwave fluctuations which should be observable by MAP and PLANCK.
The correlation between these various predictions should serve as a clear test of the B-ball baryogenesis/D-term inflation scenario. 

\subsection*{Acknowledgements}   Many thanks to Kari Enqvist
for a stimulating and continuing collaboration. Thanks also
to the European Union TMR programme, which supported most of the
work reported here, and to the Dept.
of Theoretical Physics, Helsinki
University, where it was carried out. This
report has been supported by the UK PPARC.

\end{document}